\documentclass[%
reprint,
amsmath,amssymb,
aps,
]{revtex4-2}

\usepackage{graphicx}
\usepackage{dcolumn}
\usepackage{bm}
\usepackage{makecell}

\begin{document}
	
	\preprint{APS/123-QED}
	
	\title{Utility-Scale Quantum Computation of Ground-State Energy in a 100+ Site Planar Kagome Antiferromagnet via Hamiltonian Engineering}
	
	\author{Muhammad Ahsan}
		\email{m.ahsan@uet.edu.pk}
	\affiliation{%
		Department of Mechatronics and Control Engineering,\\
		University of Engineering and Technology Lahore, Lahore 54890
	}%
	
	
	
	\date{\today}
	
	\begin{abstract}
		
	 	We present experimental quantum computation of the ground-state energy in a 103-site flat Kagome lattice under antiferromagnetic Heisenberg model (KAFH), with IBM’s Heron r1 and Heron r2 quantum processors. For spin-1/2 KAFH, our per-site ground-state energy estimate is
	 	$-0.417\,J$, which, under open-boundary corrections, matches the energy in thermodynamic limit. i.e. $-0.4386\,J$. To achieve this, we used a hybrid approach that splits the conventional Variational Quantum Eigensolver (VQE) into local (classical) and global (quantum) components for efficient hardware utilization. More importantly, we introduce a Hamiltonian engineering strategy that increases coupling on defect triangles to mimic loop-flip dynamics, allowing us to simplify the ansatz while retaining computational accuracy. Using a single-repetition, hardware-efficient ansatz, we entangle up to 103 qubits with high fidelity to determine the Hamiltonian lowest eigenvalue. This work demonstrates the scalability of VQE for frustrated 2D systems and lays the foundation for the future studies using deeper ansatz circuits and larger lattices on utility quantum processors.

	\end{abstract}
	
	\maketitle
	
	
	\section{\label{intro}Introduction}
	
	Recent advances in quantum hardware have brought general-purpose quantum processors to the threshold of solving problems that challenge classical computation, particularly in quantum chemistry and strongly correlated materials~\cite{bharti2022nisq,cao2019quantum}. Among the most promising tools is the Variational Quantum Eigensolver (VQE)~\cite{peruzzo2014variational}, a hybrid quantum-classical algorithm well-suited to current noisy intermediate-scale quantum (NISQ) devices~\cite{NISQ}, especially those following the heavy-hex architecture, such as IBM’s Falcon and Hummingbird processors~\cite{kandala2017hardware,ibm2023devices}. With public access to devices supporting over 100 qubits, VQE is being extended to simulate increasingly large systems. However, the translation of reduced noise and improved coherence times into utility-scale quantum computation remains an open question, especially in challenging domains like quantum magnetism, where ground-state preparation and energy estimation are complicated by frustration and topological order~\cite{savary2017quantum,luo2021gapless}. The central aim of this study lies in the multi-prong approach for the utility-scale quantum computing~\cite{Patra2024, Baumer2024, Sachdeva2024, Tindall2024, Chowdhury2024, ODRPaper}, a crucial milestone on the path to quantum advantage~\cite{AhsanMillionQubits}.
	
		\begin{figure}
		\includegraphics[scale = 0.43]{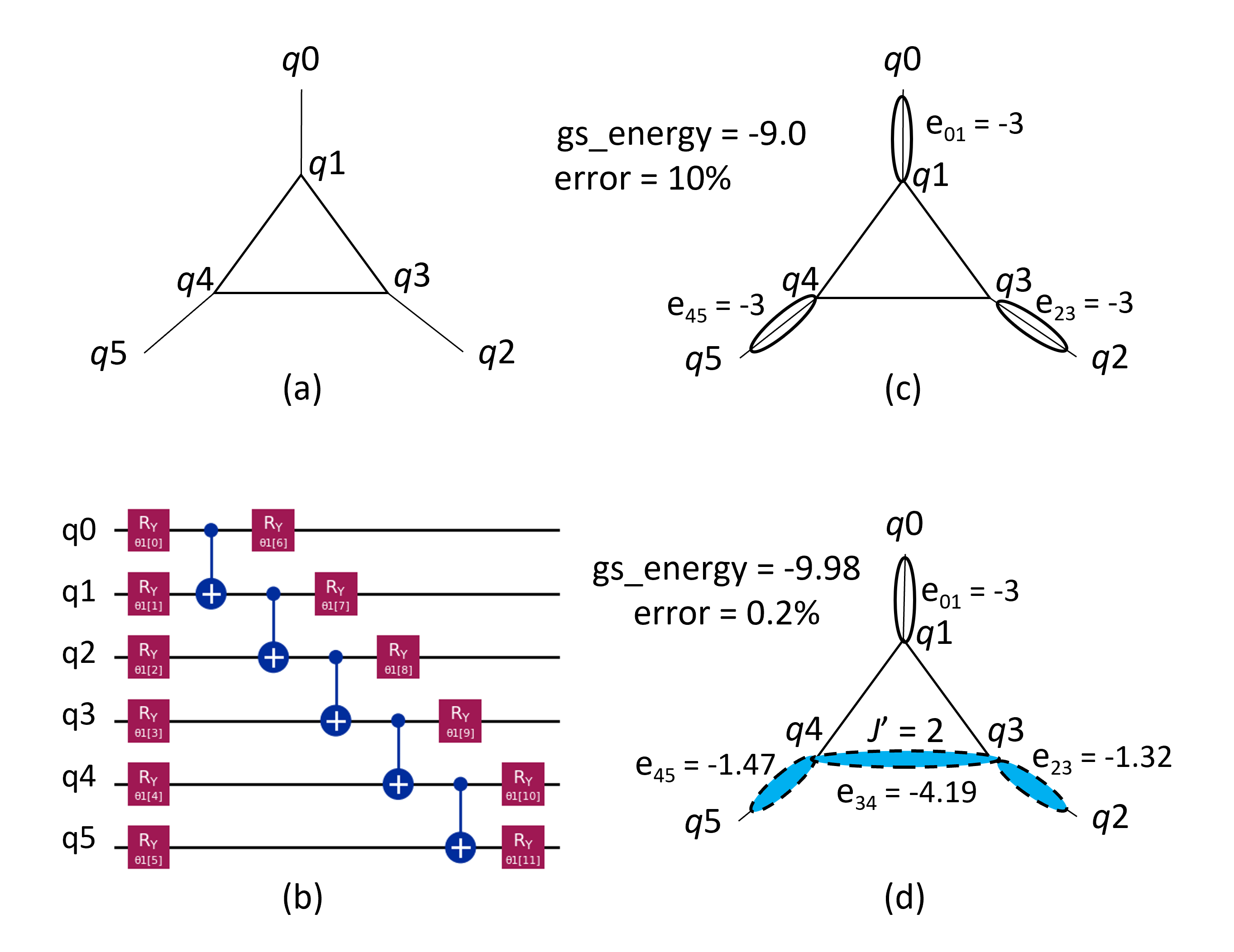}
		\caption{\label{fig:wide} 
			A proof-of-concept demonstration of Hamiltonian engineering for lowering the ground-state energy.
			(a) A 6-site subregion of the Kagome lattice targeted for ground-state energy calculation.
			(b) The corresponding 6-qubit VQE ansatz circuit containing chain of CNOT gates sandwiched between parametric $R_y(\theta)$ gates.
			(c) Ground-state energy (gs\_energy) and bond energy map $\textnormal{e}_{\texttt{XY}}$ \emph{without} Hamiltonian engineering, showing localized static dimers.
			(d) Applying Hamiltonian engineering with enhanced coupling $J' = 2$ lowers the gs\_energy. The bond energy map shows that the static dimers (q2, q3) and (q4, q5) evolve into superposition of dimers, entangling all four qubits, thus simulating the spin-liquid state.}
		
	\end{figure}
	
\begin{figure*}
		\centering
		\includegraphics[scale = 0.43]{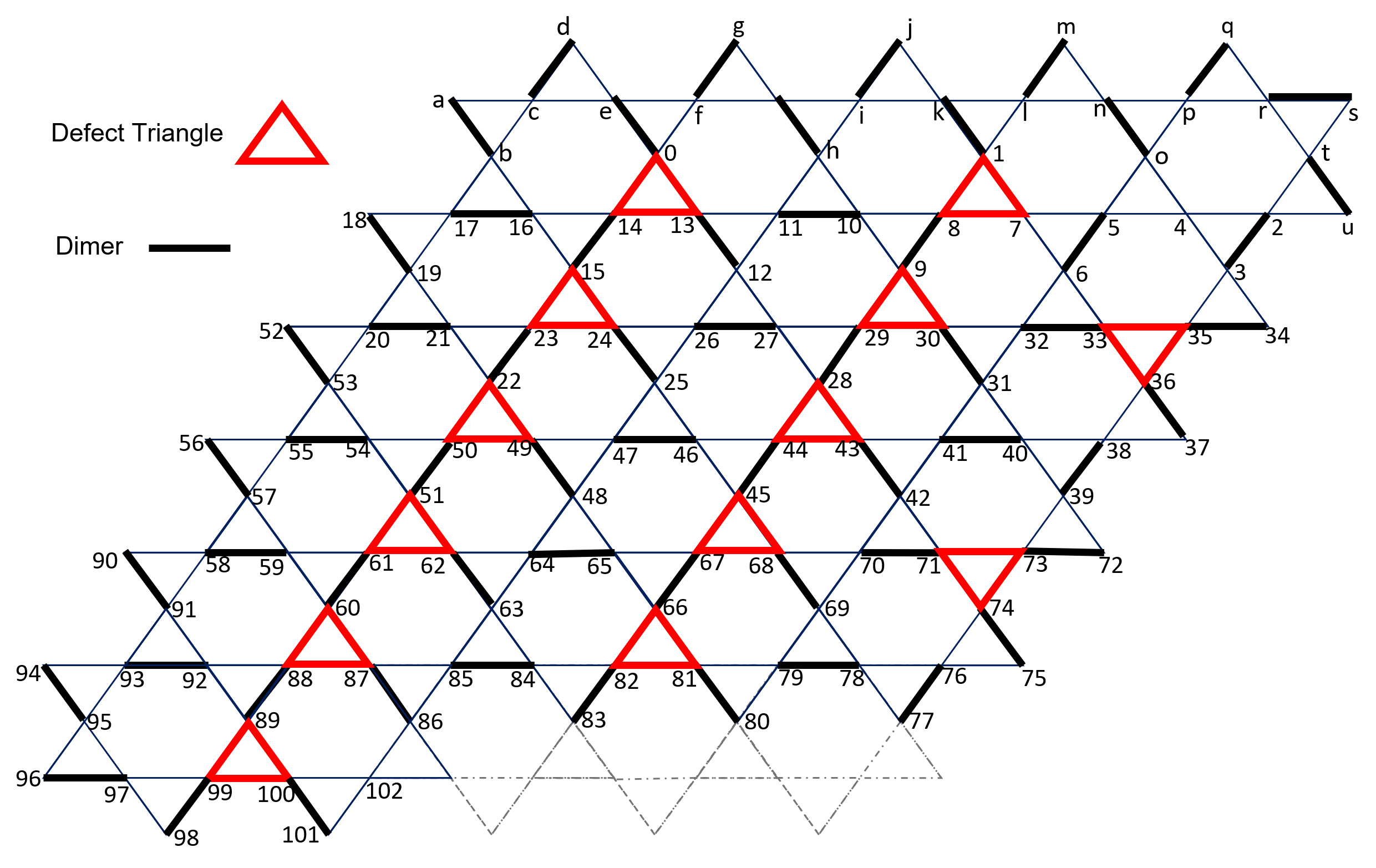}
		\caption{\label{fig:KagomeOnly} (Color online) The 125-site planar Kagome lattice showing static dimer coverings (thick black line-segments) and defect triangles (in red), driven by the our qubit-lattic site map. The numerical labels on the sites correspond to qubit indices in the 103-qubit ansatz circuit depicted in Figure~\ref{fig:AnsatzProcessor}(a). These qubits were included in the quantum computation of the gs\_energy of 103-site KAFH. Alphabetically labeled sites were excluded from the quantum circuit but were included in the final per-site ground-state energy calculation. The dotted region was omitted entirely and not mapped to the quantum problem in this study.  
		}
		
	\end{figure*}

	In this work, we showcase a scalable VQE framework to estimate the ground-state energy (gs\_energy) of the 103-site spin-1/2 Kagome Antiferromagnetic Heisenberg model (KAFH) on a fully flat 2D lattice shown in Figure-\ref{fig:KagomeOnly}. \emph{Note that throughout this manuscript, we assume exchange coupling strength $J = 1$ and, unless otherwise specified, report four times} (4x) \emph{the ground-state energy for convenience in calculations and data presentation}. This means that singlet energy is taken $-3$ instead of $-3/4$. Unlike prior studies constrained to quasi-1D geometries (e.g., cylinders or strips) for computational tractability~\cite{yan2011spin}, our method preserves the full 2D translational and topological symmetry of the lattice. We introduce a Hamiltonian engineering strategy: a valence bond crystal (VBC) dimer configuration is used as a physically motivated starting point, and local exchange couplings are modified (specifically, increasing one bond of each defect triangle to $J' \approx 2$, while keeping all others at $J = 1$) to simulate the energetic effects of loop-flips and drive dimer resonance, as highlighted in Figure-\ref{fig:wide}. This adjustment compensates the relatively shallow, hardware-efficient ansatz~\cite{kandala2017hardware} for their lower expressiveness, particularly our single-repetition real-amplitude circuit supporting only 1D nearest-neighbor entanglement, and helps induce quantum fluctuations in the vicinity of defect triangles, enabling a superposition of dimer covers reminiscent of the resonating valence bond (RVB) spin-liquid picture~\cite{anderson1987resonating,sachdev1992kagome}.

	To implement this approach efficiently, we partition our VQE algorithm into local and global phases. The local VQE is executed on classical hardware, decomposing the 103-qubit ansatz (Figure-\ref{fig:AnsatzProcessor}(a)) into overlapping 15–19 qubit subcircuits that each span defect triangles and nearby corner-sharing structures. Each subcircuit is independently optimized, with $J'$ adjusted to match known ground-state energies from exact diagonalization (feasible at these small scales). The subcircuits are then stitched into a complete ansatz with additional parametric single-qubit gates at the interfaces, see Figure-\ref{fig:vqe}. The global VQE phase runs on the quantum processor to optimize only these bridging gates, potentially capturing longer-range entanglement. Final energy estimates are refined using the Operator Decoherence Renormalization (ODR)~\cite{ODRPaper} noise-mitigation protocol~\cite{cincio2021}. This work pushes the frontier of practical quantum simulation by: (1) offloading expensive local optimization to classical hardware
	(2) enabling high-fidelity energy estimation through localized Hamiltonian engineering
	(3) demonstrating  utility-scale quantum computation 
	(4) preserving the geometric and topological integrity of the Kagome lattice during simulation.

To our knowledge, this is the first application of local Hamiltonian calibration, via bond-specific \( J \) tuning, to enhance VQE performance on large 2D Kagome lattices. In contrast to previous VQE or hybrid tensor-network approaches \cite{kattemolle2021vqe, bosse2022, javanmard2024}, which often rely on quasi-1D or complete-graph hardware, our method preserves full 2D topology and directly engineers local fluctuations crucial for RVB and topological order. We believe this technique opens a new avenue for scalable quantum simulation of frustrated magnetism and spin-liquid physics. The manuscript organization situates the prior work in Section-\ref{sec:prior}, followed by the details of hamiltonian engineering in Section-\ref{sec:hamilt_engg}. The proposed VQE algorithm is described in Section-\ref{sec:VQE} while experimental results are provided in Section-\ref{sec:exp_results}. The Section-\ref{sec:benchmarking} validates our experimentally obtained gs\_energy estimate and Section-\ref{sec:conclusion} concludes the article with possible directions for the future studies. The Qiskit code and related material is available at \url{ https://github.com/ahsan-quantum/Kagome-Lattice-VQE.git}

	\section{\label{sec:prior} Prior Work}
	
	The quest to accurately determine the ground state properties of strongly correlated quantum many-body systems remains a central challenge in condensed matter physics. Among these, geometrically frustrated magnets, such as the Kagome Antiferromagnet, are of particular interest due to their potential to host exotic quantum phases, notably the quantum spin liquid (QSL)~\cite{anderson1987resonating,balents2010spin}. The spin-1/2 Heisenberg KAFH is theoretically predicted to exhibit a gapped topological $\mathbb{Z}_2$ spin liquid ground state, characterized by robust long-range entanglement and fractionalized excitations~\cite{jiang2008accurate,depenbrock2012nature,singh2021fraction}. Classical numerical techniques like Density Matrix Renormalization Group (DMRG)~\cite{DMRG} and exact diagonalization (ED) have provided deep insights for smaller systems, but their scalability limitations restrict exploration of the thermodynamic regime. 
	VQE has emerged as a scalable alternative~\cite{kandala2017hardware,peruzzo2014variational,mcclean2016theory}, enabling ground-state estimation on NISQ devices. Early work applied VQE to 1D Heisenberg models~\cite{colless2018computation}, followed by studies of frustrated systems~\cite{guerreschi2017practical,zhang2022vqe}. Jattana et al.~\cite{jattana2022assessment} explored the 2D Heisenberg model on square and triangular lattices using VQE. More recent studies investigated frustrated lattices like the Kagome: Kattemölle and van Wezel~\cite{kattemolle2021vqe} proposed SU(2)-symmetric ans\"atze; Bosse and Montanaro~\cite{bosse2022} demonstrated $< 1\%$ energy error on small Kagome patches; Guo and Qin~\cite{qian2024} examined J1-J2 phase transitions; and Javanmard et al.~\cite{javanmard2024} implemented a hybrid MPS-VQE approach for noise-resilient circuits. Experimental implementations have also progressed: in Ref~\cite{ibmopenscience2023}, competition, winners reported sub-1\% energy error on IBM hardware, and Grossardt et al.~\cite{Weaving2025} showed contextual VQE can achieve 0.01\% accuracy for small Kagome clusters. Yet, full 2D Kagome connectivity on large quantum devices has remained largely unexplored.
	
	\section{\label{sec:hamilt_engg}Hamiltonian Engineering on the Defect Triangles}
	
	Quantum spin liquids, emergent phases characterized by long-range entanglement and a highly resonant superposition of valence bond (dimer) configurations on frustrated lattices, have attracted intense theoretical and experimental scrutiny over the past decade \cite{balents2010spin, savary2017quantum, broholm2020}. The KAFH model is a prototypical platform for realizing such physics, with compelling evidence, via DMRG and quantum dimer model methods, favoring a gapped \( \mathbb{Z}_2 \) spin liquid ground state \cite{yan2011spin, depenbrock2012nature, jiang2008accurate}. However, nearly all high-precision simulations employ quasi-1D geometries (cylinders or strips), which necessarily break the lattice’s 2D translational and topological symmetries. Here, we propose a scalable quantum computational strategy based on a 125-site planar Kagome lattice, preserving the intrinsic 2D lattice topology. 
	
	\subsection{Calibration the Hamiltonian}
	Let a $\mathcal{K}=\{(\alpha,\beta)\}$ denote set of all pairs  ($\alpha, \beta$) representing neighboring sites of the Kagome lattice shown in Figure-\ref{fig:KagomeOnly}. Accordingly the KAFH Hamiltonian $H$ becomes
	\begin{center}
		$H = J \sum_{(\alpha,\beta)\in \mathcal{K}} X_\alpha X_\beta+Y_\alpha Y_\beta+Z_\alpha Z_\beta$
	\end{center}
	and $H_{pert}$ is the engineered hamiltonian which enhances the exchange coupling $J \longrightarrow J'$ on some $\mathcal{S} \in \mathcal{K}$ to be defined on the defect triangles. Then, 
	\begin{align}\label{eq:hpert}
		H_{pert} =\;& J'_{(\alpha,\beta)} \sum_{(\alpha,\beta)\in \mathcal{S}} \left( X_\alpha X_\beta + Y_\alpha Y_\beta + Z_\alpha Z_\beta \right) \nonumber \\
		&+ J \sum_{(\alpha,\beta)\in \mathcal{K}\setminus\mathcal{S}} \left( X_\alpha X_\beta + Y_\alpha Y_\beta + Z_\alpha Z_\beta \right)
	\end{align}
	 
	\noindent Our approach begins with a static valence bond covering, a natural reference state creating one defect triangles as shown in Figure-\ref{fig:KagomeOnly}. To remediate the systematic under-representation of dimers on defect sites in the static picture, we tune one Heisenberg coupling on each defect triangle from \( J = 1 \) to \( J' \approx 2 \). This calibration aligns the energy gain from creating a new dimer with the energy penalty of removing two existing dimers, as one expects from local energy conservation arguments in RVB or dimer models. It injects loop resonances around defect triangles, encouraging the system toward spin-liquid-like superposition as showcased in Figure-\ref{fig:wide}.
	
	This localized increase in coupling aims to strongly induce dimerization within these specific triangles, making it energetically favorable for our ansatz to form a singlet. This targeted modification results in an average interaction strength of \( J_{\text{avg}} = \approx \frac{4}{3} \) on these defect triangles. By energetically penalizing non-dimerized configurations in these frustrated regions, we compel the VQE to explore quantum states which can lead VQE closer to the true Kagome ground-state energy.
	Our 125-site Kagome patch, with open boundaries in both \( X \) and \( Y \) directions, introduces additional finite-size effects and results in a higher (less negative) ground-state energy compared to the thermodynamic limit. However, the proposed Hamiltonian engineering focuses on enabling our ansatz to efficiently discover a low-energy configuration within this finite system, thereby approximating the physics of the larger Kagome lattice.
	
	The calibrated Hamiltonian approach leverages a hardware-efficient Variational Quantum Eigensolver (VQE) framework. Our ansatz is the Qiskit~\cite{qiskit2024} Real Amplitude circuit. Figure-\ref{fig:AnsatzProcessor}(a) shows that it is 1-D chain of CNOT entanglers interleaved with parameterized single-qubit rotations, designed for compatibility with IBM’s heavy-hex superconducting architecture. This circuit can entangle nearest-neighbor qubits and is sufficiently expressive to capture local resonance physics, while remaining shallow and hardware-friendly.  
	
		\begin{table}[h!]
				\caption{Effectiveness of Hamiltonian engineering with $J' \approx 2$ in matching the estimated gs\_energy \footnote{This data was obtained using Qiskit simulations} to the exact value for progressively larger subregions of the Kagome lattice, as shown in Figure-\ref{fig:SubLattices}.
			}
			\label{tab:example}
		\centering
		\begin{tabular}{|c|c|c|c|c|c|}
			\hline
			\thead{No. of\\Sites} & \thead{Subregions of \\Kagome \\ Lattice \\ of Figure-\ref{fig:SubLattices}} & \thead{No. of \\Defect \\ Triangles} & \thead{Exact \\gs\_energy\\ (units of J)} & \thead{VQE \\ computed \\ gs\_energy\\ (units of $J$) } & \thead{$J$' \\ (units \\ of \\ $J$)}\\ \hline
			6 & (a) & 1 & $-10.0$ & $-9.98 \pm 0.04$ & 2\\ \hline
			8 & (b) & 1 & $-13.02$ & $-13.03 \pm 0.07$ & 1.95\\ \hline
			12 & (c) & 3 & $-18.0$ & $-18.02 \pm 0.05$ & 2\\ \hline 
			19 & (d) & 1 & $-29.14$ & $-29.10 \pm 0.05$ & 1.9\\ \hline 
			23 & (e) & 2 & $-36.43$ & $-36.43 \pm 0.04$ & 1.95\\ \hline 
			
		\end{tabular}
	
	\end{table}
	
	\subsection{Scalability of Hamiltonian Calibration}
	We first validated the defect-triangle enhancement strategy on small Kagome lattice patches, demonstrating that in all cases, the VQE-computed energies closely matched exact diagonalization results, with relative errors arbitrarily close to zero with the fine tuning of \( J' \) around 2. This agreement confirms that selectively increasing the interaction strength on defect triangles reliably guides the ansatz towards correct ground-state energy. The success of this empirical calibration on small lattices provides strong justification for extending the same Hamiltonian engineering strategy to larger systems. Since the Heisenberg Hamiltonian is strictly local and the underlying frustration-driven physics exhibits self-similarity, we expect the tuning of \( J' \) to remain effective as system size increases.
	
	Table~\ref{tab:example} summarizes example Hamiltonian engineering for various Kagome lattice subregions containing 6 to 23 sites. For each subregion size \( N \), classical VQE simulations with \( J' \approx 2 \) yield ground-state energies in close agreement with exact diagonalization. The corresponding geometries are depicted in Figure-\ref{fig:SubLattices}, each chosen to ensure that at least one defect triangle is present and that the ansatz qubit mapping accommodates local entanglement. In some configurations (e.g., \( N=23 \)), two or more defect triangles are included. In each sublattice, the specific bond on the defect triangle where the Heisenberg coupling was approximately doubled is marked as \( J' \). The data highlights that the same calibrated interaction strength \( J' \) suffices across different lattice geometries without the need for re-tuning, reinforcing the scalability of the method.
	
	Figure~\ref{fig:SubLattices} further illustrates a representative qubit-to-lattice site mapping that defines the initial dimer cover and locate defect triangles. Although this VBC snapshot is highly dependent on the map and the ansatz circuit, it enables the broader applicability of our Hamiltonian engineering approach for accurate ground-state energy estimation in larger, general Kagome lattice geometries.
	
	\subsection{Physical Interpretation of $J' \approx 2$}
	The smallest loop in the Kagome lattice that allows dimer flipping is the hexagonal plaquette contained within a single unit cell.
	We justify setting \( J' \approx 2 \) by demonstrating that, at this value, the ansatz circuit enables a complete loop-flip and reproduces the exact ground-state energy. Figure-\ref{fig:SubLattices}(c) shows a 18-site Kagome unit cell whose gs\_energy was computed from loop version of 6-qubit hardware-efficient ansatz circuit, which applies CNOT 5,0 to complete the loop of entangling gates. The initial dimer configuration places static singlets on edges (0,1), (2,3), and (4,5), leading to an overestimated energy of gs\_energy $=-9$, while leaving three defect triangles. When the exchange coupling \( J' = 2 \) is applied on bonds (1,2), (3,4), and (5,0), the ansatz induces a loop-flip within the central hexagon and recovers the exact ground-state energy gs\_energy $=-18$. This proof-of-concept validates that targeted Hamiltonian engineering on defect triangles can correct energy errors arising from suboptimal qubit mappings or limited ansatz expressiveness.
	
		\begin{table}[h!]
		\centering
			\caption{Components of the gs\_energy computed using local VQE \footnote{This data was obtained using Qiskit simulations}. Each component corresponds to a subregion shown in Figure~\ref{fig:LocalSubLattices}, with its respective defect triangle bond and its fine-tuned $J'$.
			}
		\begin{tabular}{|c|c|c|c|}
			\hline
		    \thead{Lattice subregion \\ of the Figure-\ref{fig:SubLattices}} & \thead{Bonds with \\ $J'$} & \thead{$J$' \\ (units of $J$)} & \thead{Local VQE \\ Sub-lattice energy \\ (units of $J$)} \\ \hline
			  (a) & $(7,8),(13,14)$ & 1.9 & -27.62 \\ \hline 
			  (b) & $(23,24),(29,30)$ & 1.9 & -23.18 \\ \hline 
			  (c) & $(35,36),(43,44)$ & 1.9 & -27.38 \\ \hline 
			  (d) & $(49,50),(61,62)$ & 1.9 & -24.68 \\ \hline 
			  (e) & $(67,68),(72,73)$ & 2.0 & -22.32 \\ \hline 
			  (f) & $(81,82),(87,88)$ & 2.0 & -20.98  \\ \hline 
			  (g) & $(99,100)$ & 1.9 & -15.82 \\ \hline 
			  Overall & & & -161.98 \\ \hline
			  
		\end{tabular}
		\label{tab:LocalVQERes}
	\end{table}
	
	Although a full dimer loop-flip offers a precise route to ground-state accuracy, it becomes increasingly difficult to simulate such global flips as the Kagome lattice expands over multiple unit cells. This limitation arises due to the hardware constraints and non-local mapping of neighboring sites in the ansatz. Nevertheless, as demonstrated in Figure-\ref{fig:wide}, setting \( J' = 2 \) on a single bond of a defect triangle within a 6-site subregion can sufficient to leave more potent signatures of the spin-liquid states. With \( J' = 2 \), the same ansatz circuit enables entanglement among qubits 2, 3, 4, and 5, effectively unfreezing the static dimer cover into a superposition state that better approximates the local physics of a spin liquid. This localized resonance significantly reduces energy estimation error, even with shallow circuits and limited entanglement connectivity.

	\section{The Proposed VQE Algorithm} \label{sec:VQE}
	The Variational Quantum Eigensolver (VQE) algorithm utilizes a parameterized quantum circuit (ansatz) to prepare a quantum state on quantum hardware. It then measures the expectation value of the system's Hamiltonian and uses a classical optimizer to iteratively update the ansatz parameters to minimize the measured energy. This hybrid quantum-classical loop continues until convergence to an energy minimum, which ideally approximates the ground-state energy. Owing to its tolerance to certain types of quantum noise, VQE has emerged as one of the most viable algorithms for near-term quantum devices.
	
	\begin{figure*}
		\centering
		\includegraphics[scale = 0.575]{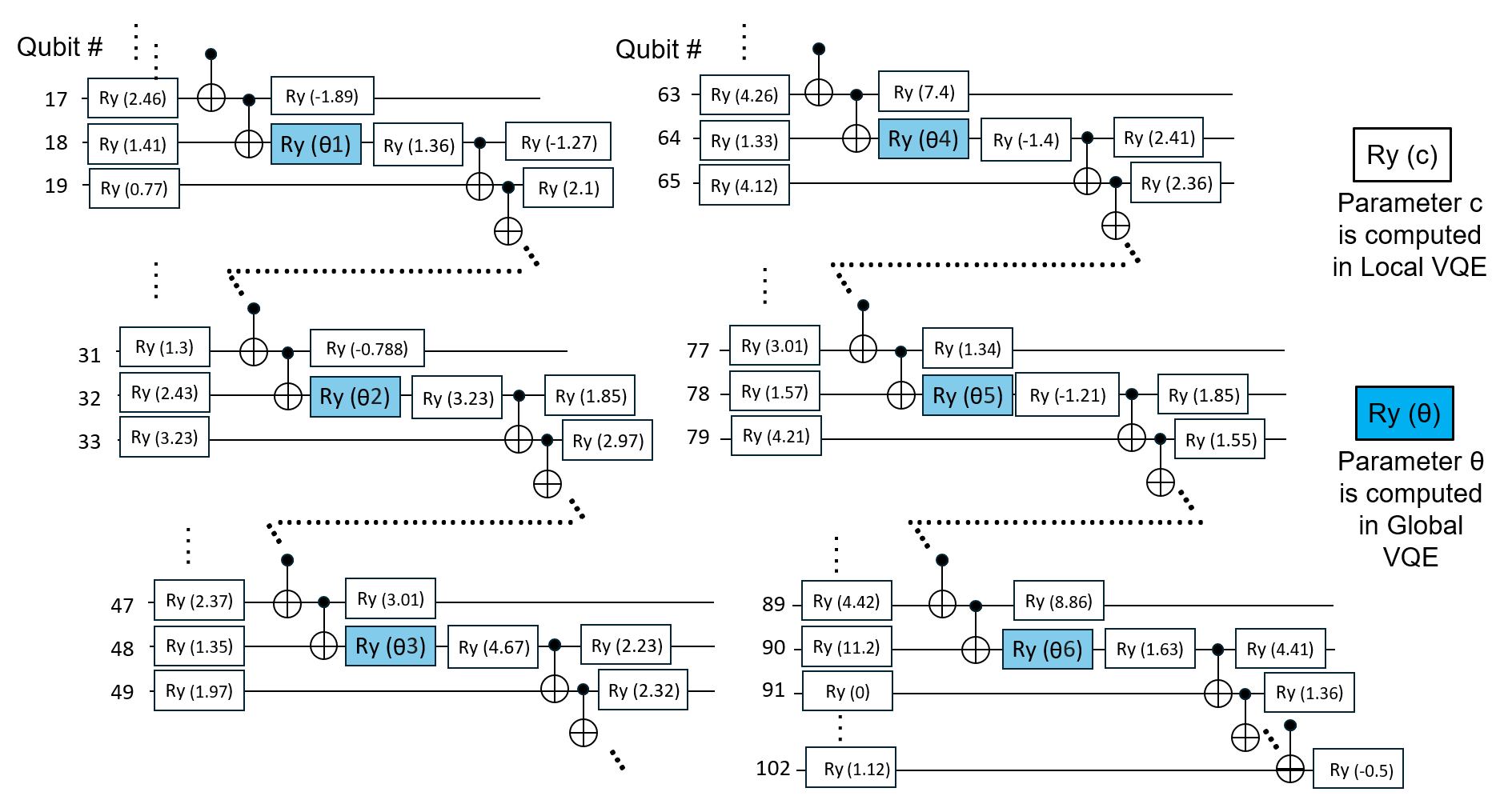}
		\caption{\label{fig:vqe} (Color online) Reconstructed ansatz circuit showing: (1) $R_y(c)$ rotation gates with precomputed angles $c$ from the local VQE, indicated by unshaded rectangles; and (2) $R_y(\theta)$ gates applied to the junction qubits, shown in shaded blue rectangles, where the angles $\theta$ are variational parameters optimized by the global VQE. The reconstructed ansatz qubit indices remain unchanged from those on the original circuit of Figure-\ref{fig:AnsatzProcessor}(a)}
		
	\end{figure*}
	
	Applying VQE to highly frustrated systems such as the KAFH model presents a unique challenge: constructing an ansatz that is expressive enough to represent complex quantum correlations while remaining efficient for execution on NISQ hardware. Hardware-efficient ansatz, composed of alternating layers of single-qubit rotations and entangling gates (e.g., CNOTs) tailored to the hardware's native topology, are particularly advantageous in this context. Our experiments employ a one-dimensional entanglement-biased ansatz, making it naturally compatible with IBM's heavy-hex superconducting architecture, which favors linear qubit connectivity.
	
	To address scalability and circuit depth limitations, we decompose the full 103-qubit VQE into two stages: local VQE and global VQE. See Figure-\ref{fig:vqe} for details. In the local phase, we optimize the parameters of smaller ansatz circuits (15--19 qubits) that correspond to subregions of the Kagome lattice displayed in Figure-\ref{fig:LocalSubLattices}. Each VQE sub-problem is independently solved on a classical simulator yielding accurate estimates of local energy contributions and the relevant rotation gate $R_y(\texttt{c})$ parameter: \texttt{c} (unshaded gates in Figure-\ref{fig:vqe}). These local solutions collectively define a baseline energy, which serves as an upper bound for the global VQE. The global phase, executed on a quantum processor, adjusts the interconnections between pre-optimized segments, allowing for long-range entanglement that can lower the total energy and approach the true ground-state of the full system.
	
	\subsection{Local VQE: Classical Computing}
	Figure-\ref{fig:LocalSubLattices} illustrates the qubit-to-site mapping and its decomposition into seven local VQE segments. Each segment covers a chain of linearly adjacent, corner-sharing triangles, and includes two defect triangles, with the exception of segment~7, which contains only one. This segmentation strategy reflects both the topological structure of the Kagome lattice and the constraints of the hardware-efficient ansatz, allowing localized parameter optimization within manageable sub-circuit sizes.
	
	A key objective of the local VQE phase is to determine a precise value of \( J' \) on selected bonds of each defect triangle. For every lattice segment, the value of \( J' \) is individually calibrated to ensure that the computed local ground-state energy matches the exact ground-state energy obtained via exact diagonalization. As shown in Table-\ref{tab:LocalVQERes}, the calibrated values of \( J' \) remain remarkably consistent across all subregions, clustering around \( J' \approx 1.95 \). This empirical regularity supports the scalability of our Hamiltonian engineering approach to larger systems. At the conclusion of the local VQE phase, we obtain a baseline estimate of the ground-state energy gs\_energy as $-161.98$ (see Table-\ref{tab:LocalVQERes}). In the subsequent phase, the sub-circuits that contributed to this value are recombined into a complete ansatz for global VQE, with the goal of further lowering the energy.

	\subsection{Global VQE: Quantum Computing}
	\subsubsection{Reconstruction of the Ansatz Circuit.}
	Local VQE produces pre-optimized ansatz segments for each subregion of the Kagome lattice. To recover the full 103-qubit circuit, these segments are concatenated such that all original gate and qubit dependencies of the ansatz are preserved, effectively reconstructing the \emph{near} original circuit. However, this naïve concatenation does not guarantee global optimality because the junction qubits, shared between two segments, inherit independently optimized \( R_y (\theta)\) rotations from each side. To address the potential sub-optimality introduced at these junctions, we insert an additional trainable \( R_y(\theta) \) gate on each junction qubit (see Figure-\ref{fig:vqe}), and tune these parameters via a global VQE.
	
	\subsubsection{Objective Function}
	Concatenating seven local ansatz segments introduces six free parameters corresponding to the newly inserted \( R_y(\theta) \) gates on the junction qubits (e.g., qubits 17, 33, 48, 63, 73, 92). The global VQE runs the reconstructed ansatz on IBM quantum processors, with the objective of minimizing \( \text{Tr}(\rho H_{\text{pert}}) \), where \( H_{\text{pert}} \) comes from eq(\ref{eq:hpert}) and $\rho$ is the quantum state prepared by the ansatz circuit. The original Hamiltonian $H$ and $H_{\text{pert}}$ include over 507 Pauli product terms and each term accumulates noise when ansatz is readout on quantum processor, thereby causing unwanted fluctuations during global VQE. To address this problem we adopt a noise-aware simplification by retaining only those terms involving the junction qubits and their immediate neighbors, this reduced Hamiltonian is denoted \( H_{\text{SEL}} \). 
	
	\[
	H_{SEL} = J\sum_{(i,j)\in \mathcal{S}} X_i X_j + Y_i Y_j + Z_i Z_j  
	\]
	where 
	
	\begin{flushleft} 
		$\mathcal{S}$ = \{(17,18),(17,19),(18,19),(30,32),(31,32),(32,33), 
		(47,48),(47,48),(48,49),(47,49),(63,64),(62,64),(64,65),\\
		(77,78),(76,78),(78,79),(89,92),(92,93),(91,92),(91,93)\}\\
	\end{flushleft}
	
	While this truncation reduces the expressiveness of the cost function, it significantly lowers the impact of noise. Owing to the entanglement path through the 102 CNOT gates of the hardware-efficient ansatz, all qubits remain technically interdependent, and thus parameter updates at a junction qubit gate $R_y(\theta)$ can, in principle, affect the quantum state of all the dependent qubits, as highlighted in the breakdown of ansatz circuit provided in Figure-\ref{fig:vqe}.
	
		\begin{figure*}
		\centering
		\includegraphics[scale = 0.575]{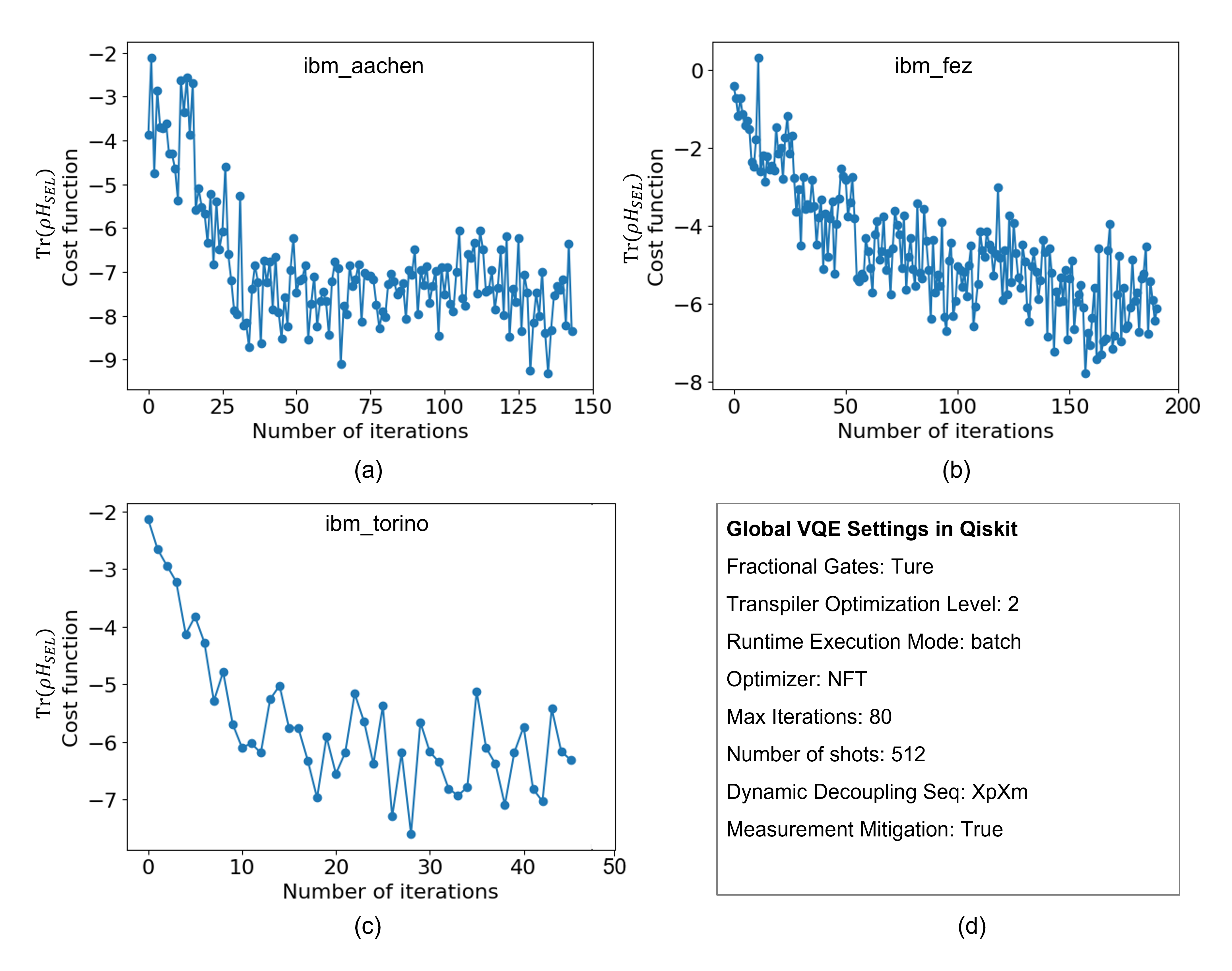}
		\caption{\label{fig:QuantVQE} Global VQE execution on the IBM quantum processors \texttt{ibm\_aachen}, \texttt{ibm\_fez}, and \texttt{ibm\_torino}, along with relevant Qiskit VQE settings. The graphs show an overall decreasing trend that stabilizes over iterations, superimposed with fluctuations driven by optimizer's algorithm~\cite{NFT}, statistics and hardware noise.
		}
	\end{figure*}
	
	\subsection{Noise Mitigation with Operator Decoherence Renormalization (ODR)}

	The ODR method~\cite{ODRPaper} is used to mitigate noise in the measurement of observables on a quantum computer. Specifically, it denoises the expectation value $\operatorname{Tr}(\rho H_{\text{pert}})$ of an observable $H_{\text{pert}}$, where the state $\rho$ is prepared by the VQE ansatz on real quantum hardware. Under the assumption of a depolarizing noise model, ODR corrects this noisy expectation value by leveraging correlations between the ideal and noisy outputs of the ansatz circuit at specific, classically tractable parameter settings.
	
	To implement this correction, we consider a parameter regime in which the VQE ansatz circuit becomes efficiently classically simulable, typically by reducing to a Clifford circuit. Importantly, the Clifford version of the ansatz arises solely due to specific parameter choices; the underlying circuit structure (i.e., the gates and their order) remains unchanged. We refer to this special configuration as the \textit{reference ansatz}.
	
	In our setting, the Clifford version of the reference ansatz prepares a static dimer cover state $|\phi\rangle$. On ideal (noiseless) quantum computer the circuit yields the expectation value $\langle\phi|H|\phi\rangle = -147$. On the other hand, when the reference ansatz circuit is executed on the quantum device, it produces a noisy (mixed) state $\sigma$, and the expectation
	$\operatorname{Tr}(\sigma H)$. In contrast, the original ansatz of the VQE algorithm typically operates in a non-Clifford regime, as it explores the full parameter space to minimize energy. Nevertheless, once the VQE optimization is complete, the ODR technique can be applied \textit{post hoc} to mitigate the noise in the final gs\_energy estimate.
	
	According to the ODR formalism, the noise-mitigated expectation value of $H_{\text{pert}}$ for the final VQE state $\rho$ is given by: 
	
	\begin{center}
		$\frac{\text{noise-mitigated} \operatorname{Tr}(\rho H_{\text{pert}})}{\operatorname{Tr}(\rho H_{\text{pert}})} = \frac{\langle{\phi}|H|\phi\rangle}{\operatorname{Tr}(\sigma H)}$
	\end{center}
	
	or equivalently,
	\begin{equation} \label{eq:2}
		\text{noise-mitigated}\operatorname{Tr}(\rho H_{\text{pert}}) = \frac{\langle{\phi}|H|\phi\rangle}{\operatorname{Tr}(\sigma H)}\cdot \operatorname{Tr}(\rho H_{\text{pert}})
	\end{equation}
	
	\noindent This correction is applied only after the VQE algorithm has converged to its optimal parameters. The technique enables partial noise mitigation by anchoring the noisy measurement against a known reference state prepared under identical circuit structure, thus enhancing the reliability of the final ground-state energy estimate obtained on noisy intermediate-scale quantum (NISQ) devices.
	
	\section{Experimental Results}\label{sec:exp_results}
	Empirical results confirm that minimizing \( \text{Tr}(\rho H_{\text{SEL}}) \) more reliably reveals downward trends in ground-state energy, compared to using the full \( H_{\text{pert}} \). Figure-\ref{fig:QuantVQE} presents objective function trajectories across iterations on IBM’s \texttt{ibm\_fez} and \texttt{ibm\_aachen} and \texttt{ibm\_torino}, with optimizer metadata, circuit depth, and noise mitigation settings. The curves exhibit an overall decrease in energy superimposed with fluctuations, primarily due to rapid optimizer updates (we used NFT~\cite{NFT}), statistical noise, and hardware errors. Among several minima points found during global VQE, we selected parameter sets achieving minimal objective function values and post-processed them with ODR to full noise-mitigated \( \text{Tr}(\rho H_{\text{pert}}) \). 
	
	Table-\ref{tab:list_opt_param} presents the optimal parameter sets obtained from the global VQE runs on ibm\_aachen, ibm\_fez, and ibm\_torino. These optimal parameters appear to be significantly influenced by processor-specific factors, including calibration quality and error rates. Nevertheless, there is no fundamental limitation preventing cross-testing of these parameter sets across different processors, which can help expand the search space for identifying the lowest gs\_energy estimate. Therefore, treating Table-\ref{tab:list_opt_param} as a general candidate pool, we proceed to compute the noise-mitigated $\textnormal{Tr}(\rho H_{pert})$ by trial assignment of each parameter set to the blue-shaded gates in Figure-\ref{fig:vqe}, across all processors.

	The values reported in Table-\ref{tab:final_result} were obtained by selecting, for each processor, the parameter set from Table-\ref{tab:list_opt_param} that yields the minimum noise-mitigated $\textnormal{Tr}(\rho H_{pert})$. To ensure consistency, we aim to maximize the ratio $\frac{\textnormal{Tr}(\rho H_{pert})} {\textnormal{Tr}(\sigma H)}$ within the same Qiskit \texttt{job}. This strategy avoids device-level behavioral and calibration fluctuations that can occur when measurements are spread across multiple Qiskit \texttt{estimator} jobs. The numerical values in Table-\ref{tab:final_result} are based on at least three independent repetitions of the same configuration (circuit, parameter set, and estimator settings). See Figure-\ref{fig:vqe}(d) for details on the estimator settings. Table-\ref{tab:final_result} summarizes energy estimates from five different quantum processors, fall between –165.91 and –172.32. 

	For comparison, classical simulations using Qiskit's Aer backend with the Matrix Product State (MPS) method and circuit cutting~\cite{CktCutting} yield an estimate –167.95, closely matching the quantum version of the results. This validates the accuracy of our noise-mitigated global VQE. To benchmark our results, we introduce heuristic energy bounds. The 103-site system contains 60 triangles, each contributing –3 in the ideal dimerized configuration, giving a theoretical lower bound of –180. However, defect triangles systematically raise the energy. For example, a single defect triangle on the smallest sub-lattice (two-unit cells as in Figure-\ref{fig:SubLattices}(d)) elevates gs\_energy by 0.86 (Table-\ref{tab:example}(d)). With 13 such defects, and each contributing $\approx$+0.86\, the effective upper bound becomes –168.82. This value aligns closely with our best estimates from both quantum and classical simulations, reinforcing the accuracy of our Hamiltonian-engineered VQE strategy.
	
		\begin{table}[h!]
		\centering
			\caption{The unmitigated (Actual) and noise-mitigated (Noise mitigated) gs\_energy using ODR eq(\ref{eq:2}), for five different ibm quantum processors}
		\begin{tabular}{|c|c|c|c|}
			\hline
			\thead{Processor} & \thead{Actual \\$\textnormal{Tr}(\rho H_{pert})$ \\ (units of $J$)} & \thead{Ref. ODR \\$\textnormal{Tr}(\sigma H)$ \\ (units of $J$)} & \thead{Noise mitigated \\
			$\textnormal{Tr}(\rho H_{pert})$ \\ (units of $J$) } \\ \hline
			ibm\_torino & $-122.81 \pm 0.53$  & $-107.28 \pm 1.68$ & $-168.28$\\   \hline
			ibm\_fez &  $-113.67 \pm 1.48$  & $-96.97 \pm 2.05$ & $-172.32$\\   \hline
			ibm\_kingston & $-122.54 \pm 1.58$  & $-108 \pm 1.15$ & $-165.91$\\   \hline
			ibm\_aachen & $-132.97 \pm 1.36$  & $-114 \pm 0.35$ & $-170.12$\\   \hline
			ibm\_marrakesh & $-115.12 \pm 0.24$  & $-100.23 \pm 2.12$ & $-168.84$\\   \hline
		\end{tabular}
		\label{tab:final_result}
	\end{table}

	\subsection{Quantum Validation}

	From a complementary perspective, this study introduces the concept of \textit{quantum validation}. The Qiskit Aer matrix product state (MPS) simulator can simulate quantum circuits with up to 64 qubits by approximating the global quantum state as a tensor network composed of locally entangled states. This approach is particularly effective for representing quantum states with weak non-local entanglement and offers a compact, efficient representation in such regimes.
	
	When applied to the local VQE framework, however, the MPS simulator inherently approximates the full quantum state and neglects the contribution of weak long-range entanglement to the ground-state energy. In contrast, global VQE executed on quantum hardware operates directly on the un-approximated quantum state and can therefore serve as a validation mechanism for the baseline ground-state energy estimate produced by the local VQE.
	
	\begin{figure*}
		\centering
		\includegraphics[scale = 0.52]{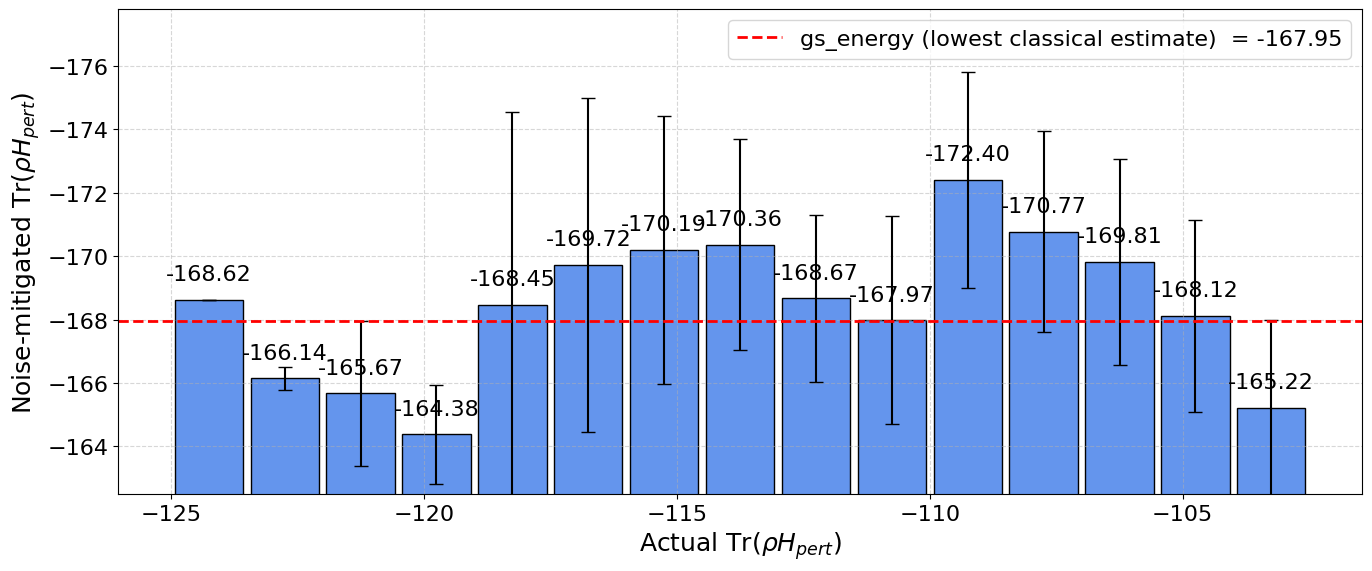}
		\caption{\label{fig:BarComp} Bar chart showing a statistical comparison of the gs\_energy computed using global VQE parameters (see Table~\ref{tab:list_opt_param}), shown as bars, and the gs\_energy computed using the parameter set $\texttt{p}_1$, obtained from the Qiskit VQE with Aer MPS simulator and circuit cutting. Error bars represent one standard deviation. 
		The chart is constructed by clustering the actual (unmitigated) $\textnormal{Tr}(\rho H_{pert})$ values into bins of size 1.5, which are then plotted against the noise-mitigated $\textnormal{Tr}(\rho H_{pert})$, \textbf{averaged} over all processors and parameter sets in Table~\ref{tab:list_opt_param}. 
		This analysis of global VQE indicates that, considering noise margins, the value \(-172.4\) is the most reliable estimate of the lowest ground-state energy, which is lower than its classical counterpart of \(-167.95\).}
		
	\end{figure*}
	
	\subsection{Comparison of Classically-optimized and Quantum-optimized Parameters}
	To further evaluate the effectiveness of parameter optimization strategies, we compared two sets of rotation angles for the junction-qubit \( R_y(\theta) \) gates. The first set, \( \mathbf{p}_1 = [5.03,\ -0.22,\ 11.65,\ 5.21,\ 4.48,\ 2.09] \), was obtained via circuit cutting~\cite{CktCutting} and classical VQE using Qiskit’s Aer backend with Matrix Product State (MPS) simulator capable of simulating up to 64 qubits. The second set, \( \mathbf{p}_2 = [5.34,\ 6.39,\ -1.24,\ 5.52,\ -2.31,\ 2.03] \) for ibm\_torino, ibm\_marrakesh, ibm\_fez, while \( \mathbf{p}_2 = [5.38,\ 0.72,\ -0.80,\ -1.42,\ 4.23,\ 1.58] \) for ibm\_kingston, was derived entirely from quantum VQE runs on real hardware by selecting the best-performing parameter set that minimized the energy objective.
	
To ensure a fair comparison, both parameter sets were independently evaluated on three quantum processors, \texttt{ibm\_torino}, \texttt{ibm\_marrakesh}, and \texttt{ibm\_fez}, by plugging them into the full 103-qubit ansatz and measuring the ground-state energy \textit{without} applying any noise mitigation (e.g., ODR). The unmitigated energy values obtained from \( \mathbf{p}_1 \) were $-121.47 \pm 1.05$, $-105.58 \pm 0.72$, and $-115.77 \pm 0.46$, respectively. In contrast, the energy values using \( \mathbf{p}_2 \) were consistently lower: $-122.81 \pm 0.53$, $-106.63 \pm 0.33$, and $-117.32 \pm 0.22$. Although the numerical differences are moderate, these results suggest that parameters learned from quantum VQE directly on hardware not only reflect the underlying circuit structure but also adaptively account for calibration specifics and coherent noise characteristics~\cite{Ahsan22}, potentially navigating toward lower-energy configurations. This finding supports the view that quantum-optimized parameters can outperform those obtained from classical approximations, particularly when hardware-specific factors significantly influence algorithm convergence.

	\begin{table}
		\centering
		\caption{Comparison of unmitigated ground-state energies using parameters from classical VQE (\( \mathbf{p}_1 \)) and global (quantum) VQE (\( \mathbf{p}_2 \)) on different IBM quantum processors.}
		
		\label{tab:param_comparison}
		\begin{tabular}{|l|c|c|c|}
			\hline
			\thead{\textbf{Processor}} & \thead{\textbf{Energy with} \( \mathbf{p}_1 \) \\ (units of $J$)} & \thead{\textbf{Energy with} \( \mathbf{p}_2 \)\\ (units of $J$)} & \thead{\textbf{Average} \\ \textbf{Energy} \\ \textbf{Decrease} \\(units of $J$)} \\ \hline
			\texttt{ibm\_torino}     & –121.47 ± 1.05 & –122.81 ± 0.53 & 1.34 \\ \hline
			\texttt{ibm\_marrakesh}  & –105.58 ± 0.72 & –106.63 ± 0.33 & 1.05 \\ \hline
			\texttt{ibm\_fez}        & –115.77 ± 0.46 & –117.32 ± 0.22 & 1.55 \\ \hline
			\texttt{ibm\_kingston}        & –123.78 ± 1.03 & –124.78 ± 0.93 & 1.00 \\ \hline
		\end{tabular}
	\end{table}
	
	An important objective of this study is to demonstrate utility-scale quantum computation. We have already shown that the parameter set $\texttt{p}_2$, obtained via quantum execution, slightly outperforms the classical counterpart $\texttt{p}_1$ in one-to-one comparisons on specific processors. Figure-\ref{fig:BarComp} presents a more holistic, statistically informed comparison of global VQE parameter performance on quantum processors with that of $\texttt{p}_1$ on classical computer. The figure contrasts the quantum gs\_energy estimates, averaged over all parameter sets listed in Table-\ref{tab:list_opt_param} and all five processors, with the classical baseline estimate using $\texttt{p}_1$ (i.e., $-167.95$).
	
	The bar chart specifically compares noise-mitigated values of $\textnormal{Tr}(\rho H_{pert})$ with the actual (unmitigated) values, which are grouped into bins of size 1.5. Notably, 7 out of 10 bars have a mean value lower than $-167.95$; however, due to large variances, these cases do not definitively outperform the classical benchmark. Still, two values, $-168.62$ and $-172.4$, stand out as statistically significant instances where the quantum estimates remain lower than the classical value, even when accounting for their upper error bounds. These values were obtained using ibm\_torino and ibm\_fez, respectively. Among them, the latter sets the lowest and most reliable gs\_energy estimate, which will be used for benchmarking per-site gs\_energy in the subsequent discussion.
	
	As quantum hardware continues to improve in gate fidelity and qubit coherence times, quantum validation of classically approximated quantum algorithms opens an exciting avenue for future research. It provides a concrete benchmark for assessing the reliability of classical simulations, especially as quantum processors scale to larger qubit counts with increasing gate fidelity. Our work presents an initial step toward this promising direction and encourages further exploration of its potential.
	
	\section{Benchmarking Against the Thermodynamic Ground State Energy} \label{sec:benchmarking}
		
		The exact ground-state energy of the spin-$\frac{1}{2}$ kagome antiferromagnetic Heisenberg (KAFH) model in the thermodynamic limit remains a cornerstone benchmark in quantum magnetism. High-precision density matrix renormalization group (DMRG) calculations have placed this value near $-0.4386(5)J$ per site~\cite{depenbrock2012nature}, a result corroborated by earlier series expansions and independent DMRG studies~\cite{ZengMarston1995, jiang2008accurate}. While theoretical frameworks for implementing variational quantum eigensolvers (VQE) on frustrated spin systems have been proposed~\cite{Wang2025}, practical demonstrations of VQE on large, two-dimensional frustrated lattices such as the KAFH remain scarce.
		
		In this work, we present what is, to our knowledge, the first experimental attempt to estimate the ground-state energy of a 103-site KAFH lattice on superconducting quantum hardware. By employing a Hamiltonian engineering strategy, we modulate bond strengths to favor valence bond configurations and optimize energy estimation using a hardware-efficient VQE ansatz. On IBM quantum processors, our estimate for the ground-state energy per site of a 103-site KAFH lattice requires conversion to the unscaled version of gs\_energy becomes $-172.4J/4 = -52.1J$ translating into per-site gs\_energy $=-521.J/103 = \texttt{-0.417}J$. This result lies above the thermodynamic-limit energy, which has been determined via large-scale DMRG and tensor network methods to lie between $-0.436$ and $-0.438J$~\cite{depenbrock2012nature, jiang2008accurate}.
		
		A substantial part of this discrepancy arises from finite-size and boundary effects, which are well known to raise energy estimates in systems with open boundary conditions. In our lattice geometry, 25 out of 125 sites lie on the boundary and possess only two nearest neighbors, as opposed to the four neighbors found in the bulk. This reduced local coordination diminishes quantum fluctuations and suppresses dimer resonance, leading to locally elevated bond energies and an increased average energy per site.
		
		
		\subsection{Open Boundary Correction}
		To account for these edge effects, we apply an open-boundary correction (OBC) model, following established procedures from prior finite-size studies~\cite{ZengMarston1995, singh2021fraction}. 
		We decompose the total energy of 125-site lattice into contributions from bulk and edge sites. The bulk sites have all four neighboring sites while edge sites lie on the boundary and have strictly fewer neighbors (usually 2).  Following the approach used in prior DMRG and tensor network studies~\cite{yan2011spin,jiang2008accurate,Shibata2011}, we express energy as weighted sum of its bulk sites' and edge sites' energies as:
		
		\begin{equation}
			E_{\text{total}} = 
			\bar{E}_{\text{bulk}} N_{\text{bulk}}
			+ \bar{E}_{\text{edge}} N_{\text{edge}} \nonumber
		\end{equation}
		
		\noindent Where $N_{\text{bulk}}$ and $\bar{E}_{\text{bulk}}$ represent number and per-site energy of bulk sites while $N_{\text{edge}}$ and $\bar{E}_{\text{edge}}$  represent number and per-site energy of the edge sites. The corrected gs\_energy, defined as $\bar{E}_{\text{bulk}}$, is given by following expression:
		
		\begin{equation}\label{eq:OBC}
			\bar{E}_{\text{bulk}} =  \frac{E_{\text{total}} - N_{\text{edge}} \cdot \bar{E}_{\text{edge}}}{N_{\text{bulk}}} 
		\end{equation}
		
		Note that $\bar{E}_{\text{bulk}}$ and $\bar{E}_{\text{edge}}$ are obtained from the unscaled version of gs\_energy divided by the number of sites. To use the eq(\ref{eq:OBC}), revert gs\_energy to the un-scaled version and estimate the average edge-site energy $\bar{E}_{\text{edge}}$ using small lattice benchmarks, where exact diagonalization is possible for given $E_{\text{total}}$. Thus, for a 12-site Kagome unit cell, $E_{\text{total}}$ = $-18J/4 = -4.5J$. Likewise, for the 19-site two-unit-cell system, we have $E_{\text{total}} = -29.14J/4 = -7.285J$, and for the 23-site case (one site less than three full unit cells), $E_{\text{total}} = -36.43J/4 = -9.1075J$. 
		For each of these cases, eq(\ref{eq:OBC}) yields the corresponding edge-site energy values: $\bar{E}_{\text{edge}} = -0.3114J$, $-0.3076J$, and $-0.3296J$, respectively. 
		Therefore, we estimate that $\bar{E}_{\text{edge}}$ lies within the range:
		
		\[
		\bar{E}_{\text{edge}} \in [-0.3076J,\,-0.3296J]
		\]
		
		\noindent For 125-site lattice, we calculate $\bar{E}_\text{bulk}$ minimum and maximum per-site energies. For a 125-site lattice our gs\_energy becomes $-172.5-36 = -208.4J$, which gives $E_{\text{total}} = 52.1J$. The Figure-\ref{fig:KagomeOnly} shows that $N_{\text{edge}} = 25$ , $N_{\text{bulk}} = 100$. For the range $[-0.3076J,-0.3296J]$, eq(\ref{eq:2})  implies 
		
		\begin{align}
		\bar{E}_{\text{bulk}}(\text{max}) = & -0.4386J \nonumber \\
		\bar{E}_{\text{bulk}}(\text{min}) = & -0.4441J \nonumber 
		\end{align}
		
		\noindent The bulk energy values reflect the underlying quantum correlations more accurately than the raw per-site average and closely match the one obtained in the thermodynamic limit  $-0.4386J$~\cite{depenbrock2012nature,Nishimoto2013},
				
		This analysis reinforces the potential for scaling our method to larger kagome lattices, where the proportion of boundary sites diminishes and finite-size effects are further suppressed. In such bulk-dominated systems, the accuracy of VQE-derived ground-state energy estimates is expected to improve, strengthening the case for quantum simulation of frustrated spin systems using near-term quantum devices.

	\section{Conclusion} \label{sec:conclusion}
	
	The methods and experimental results presented in this study demonstrate a scalable and physically grounded approach for estimating the ground-state energy of the spin-$\frac{1}{2}$ kagome antiferromagnetic Heisenberg model on large general Kagome lattices. By employing Hamiltonian engineering, specifically, increasing the exchange coupling strength on selected bonds within defect triangles, we circumvent several limitations commonly encountered in prior studies. These include constraints on system size, dimensionality, circuit depth, and the expressibility of the variational quantum eigensolver ansatz. Importantly, our approach alleviates the need for preparing quantum states with complex, long-range entanglement, which remains a significant challenge for current hardware architectures.
	
	Our use of enhanced exchange coupling is not only a computational convenience, but also deliberate strategy that leverages core features of quantum spin liquids, such as resonating valence bond (RVB) physics, loop flips, and the quantum superposition of multiple dimer coverings. By initializing the VQE with a physically motivated valence bond crystal (VBC) configuration, we guide the optimizer toward low-energy regions of the Hilbert space while maintaining hardware efficiency. This hybrid strategy, melding insights from many-body physics with the practical constraints of NISQ devices, enables an accurate and resource-conscious estimation of the ground-state energy on systems as large as 125 sites.
	
	A natural extension of this work can be scaling to larger kagome lattices and exploring Hamiltonian engineering schemes that adapt dynamically based on lattice geometry and local entanglement structure. Investigating how different defect patterns or spatially varying couplings influence dimer dynamics and entanglement entropy could deepen our understanding of emergent phases in frustrated magnets. Also, it will be interesting to explore the impact of multiple repetition hardware efficient or other more expressive ansatz for the preparation of quantum state closer to the spin-liquid. As quantum hardware continues to evolve, incorporating mid-circuit measurements, qubit resets, or adaptive feedback into VQE protocols may allow more expressive ansatz and more accurate energy estimates. In the long term, coupling these techniques with tensor network hybridizations or error-mitigation protocols could help bridge the gap between NISQ-era capabilities and fully fault-tolerant quantum simulation of strongly correlated quantum matter. 
	
	We provide our implementation as open-source to support transparency and quantum computing community engagement at \url{https://github.com/ahsan-quantum/Kagome-Lattice-VQE.git}. 
	
	\begin{acknowledgments}
		We acknowledge the use of IBM Quantum services, including IBM Quantum Credits for this work. The views expressed are those of the authors, and do not reflect the official policy or position of IBM or the IBM
		Quantum team. Our experiments used following IBM quantum processors: \textbf{ibm\_aachen}, \textbf{ibm\_kingston}, \textbf{ibm\_torino}, \textbf{ibm\_fez} and\textbf{ ibm\_marrakesh}. We also acknowledge the support of Higher Education Commission Pakistan under the project National Center of Quantum Computing.
		
	\end{acknowledgments}


\appendix
\section{Supplementary material}
\begin{figure*}
	\centering
	\includegraphics[scale = 0.45]{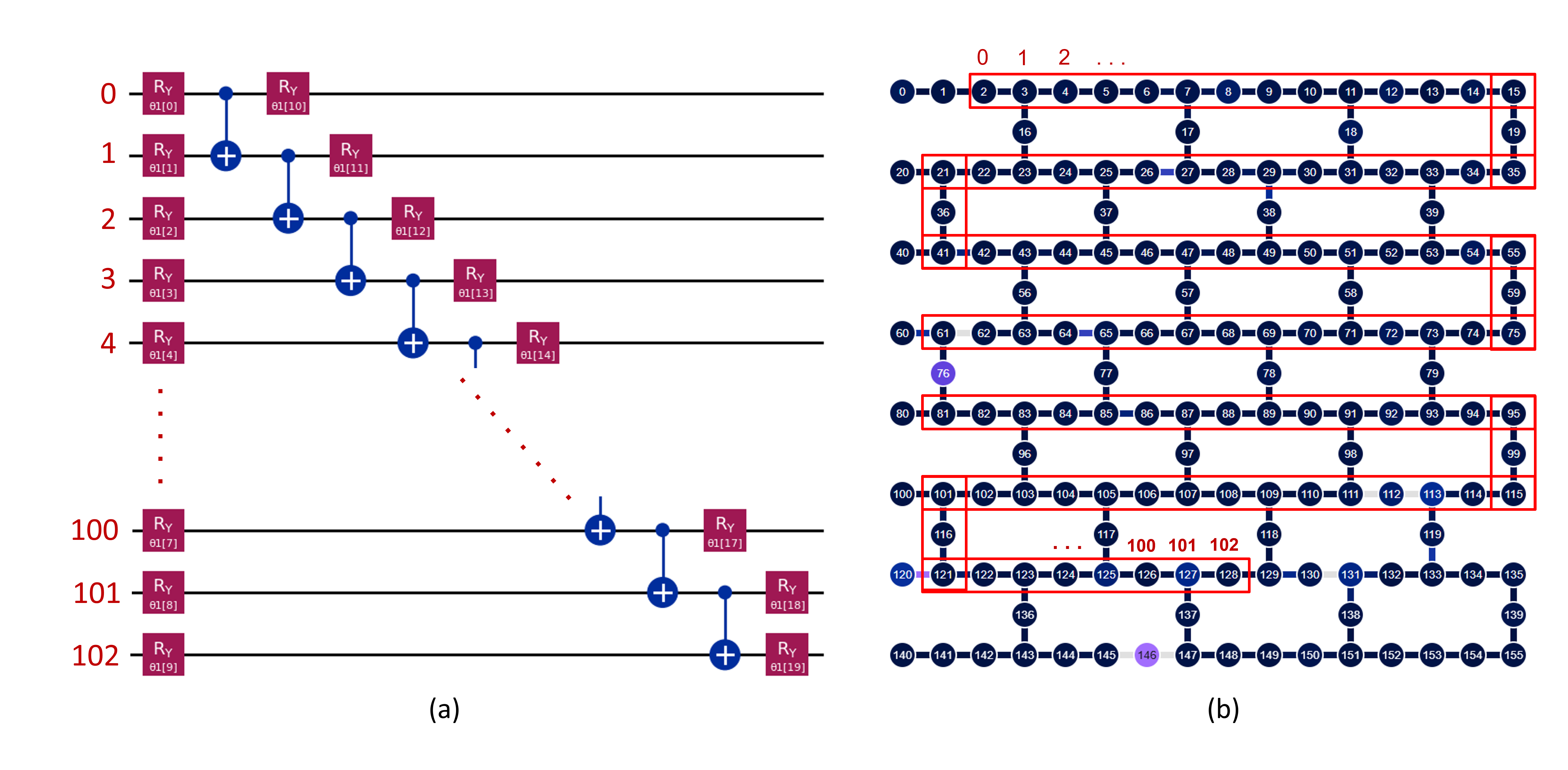}
	\caption{\label{fig:AnsatzProcessor} (Color online) (a) VQE ansatz circuit with qubit indices mapped to the corresponding lattice sites on the Kagome lattice shown in Figure~\ref{fig:KagomeOnly}. 
		(b) Physical qubit layout of the IBM Heron r1 processor (\texttt{ibm\_kingston}), showing how circuit qubits are assigned to hardware qubits. 
		The complete mapping can be followed by tracing the red outline from qubit 0 through 1, 2, $\ldots$, up to 100, 101, and 102.}
	\end{figure*}
	
	\begin{figure*}
	\centering
	\includegraphics[scale = 0.5]{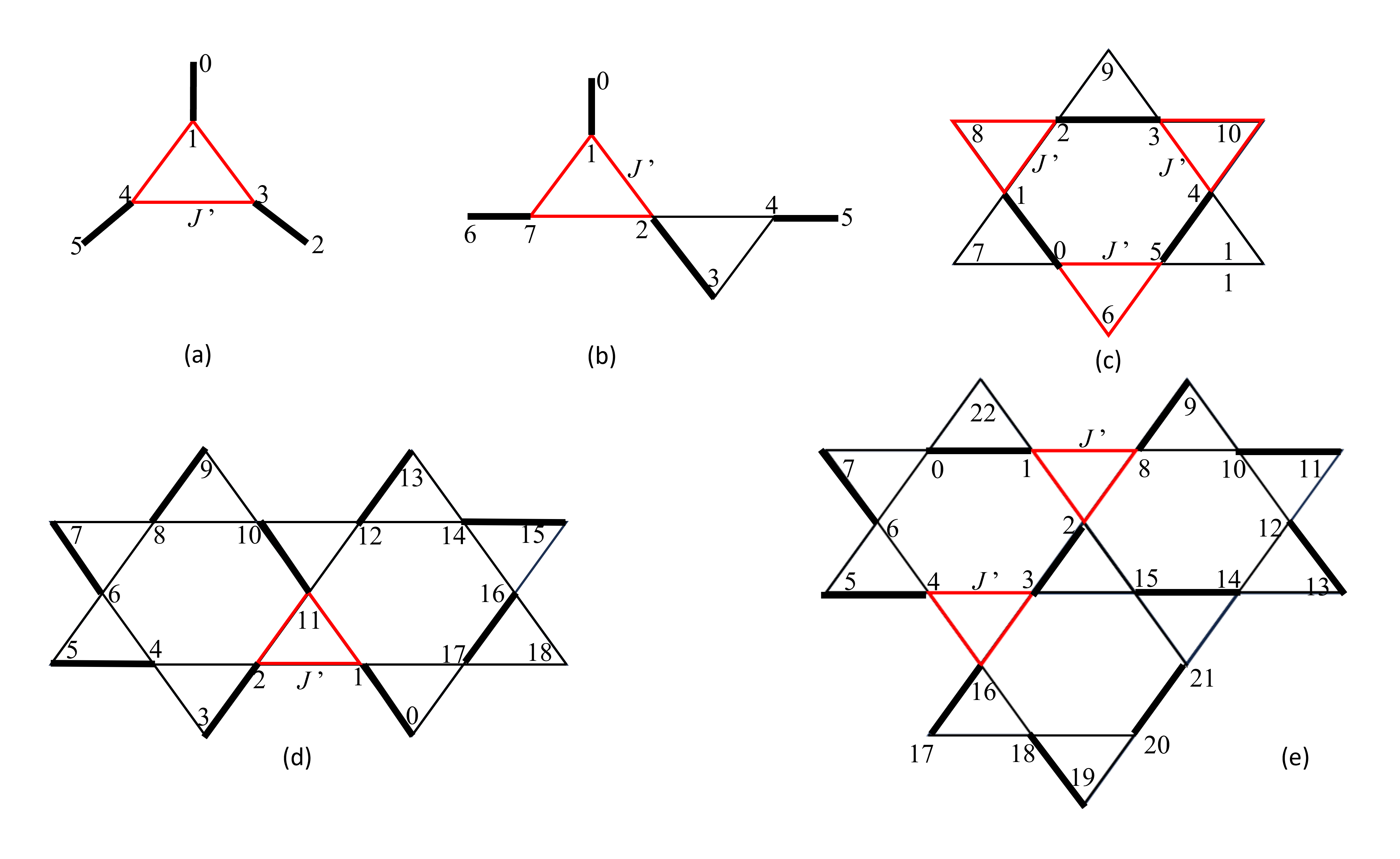}
	\caption{\label{fig:SubLattices} (Color online) Example subregions of the Kagome lattice used to demonstrate the effectiveness of Hamiltonian engineering, as discussed in Table~\ref{tab:example}. Numerical labels on the lattice sites correspond to qubit indices in the associated ansatz circuits (not shown), which are scaled versions of the circuit in Figure~\ref{fig:wide}(b). 
	The qubit mapping defines a specific dimer covering (thick black line-segments) and identifies the defect triangle (in red). 
	Bonds on the red triangles with enhanced coupling strength are labeled as $J'$.}

	\end{figure*}

	\begin{figure*}
		\centering
		\includegraphics[scale = 0.4]{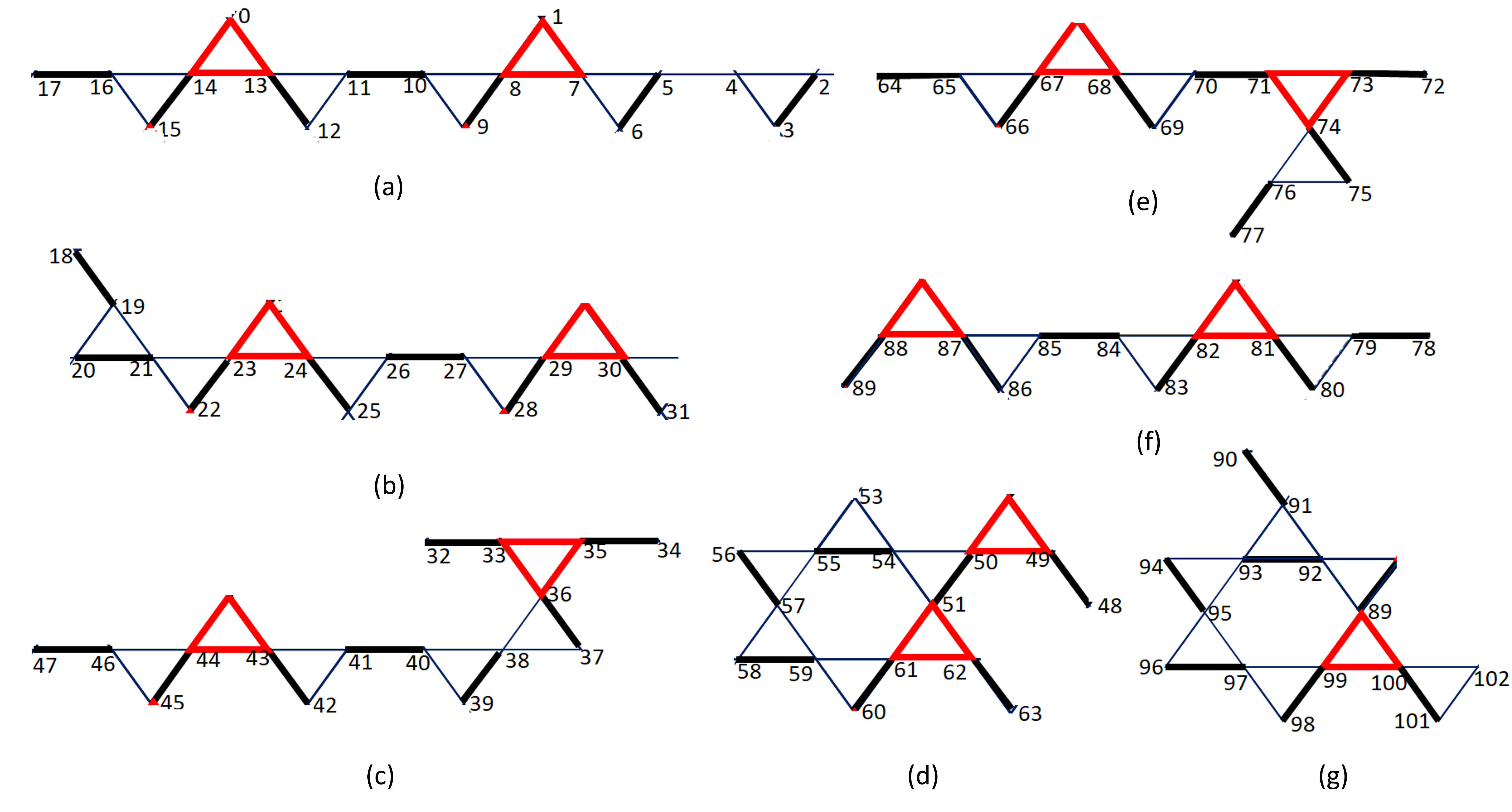}
		\caption{\label{fig:LocalSubLattices} (Color online) Subregions of the Kagome lattice of Figure-\ref{fig:KagomeOnly} used in local VQE, with corresponding ground-state energy components listed in Table~\ref{tab:LocalVQERes}. 
		Numerical labels on the lattice sites indicate qubit indices from the ansatz circuit shown in Figure~\ref{fig:AnsatzProcessor}(a). 
		The qubit mapping defines a specific dimer covering (thick black line segments), and highlights the defect triangle (in red). 
		Table~\ref{tab:LocalVQERes} also lists the bonds selected for Hamiltonian engineering within each defect triangle, along with their enhanced coupling values $J'$.}

	\end{figure*}

\begin{table} [h!] 
	\centering
	\caption{Optimal parameters computed using global VQE for different IBM quantum processors. These are obtained from the experiments data of Figure-\ref{fig:QuantVQE}.}
	
	\begin{tabular}{|c| c c c c c c|}
		\hline
		Processor &  $\theta_1$ & $\theta_2$ & $\theta_3$ & $\theta_4$ & $\theta_5$ & $\theta_6$ \\ \hline
		
		& [5.05 & -0.33 & 6.94 & -1.56 & 4.70 & 1.40] \\ 
		\makecell{\texttt{ibm\_aachen}} & [-1.78 & -0.73 &  5.24 & 4.94 & -1.66 & 1.20] \\ 
		& [5.34 & -0.24 &  3.20 & -0.18 & 3.49 &  1.35] \\
		\hline
		
		& [5.39 &  5.75 & -0.72 &  5.01 &  -1.64 & 1.06] \\
		\makecell{\texttt{ibm\_fez}} & [-1.77 & -0.73 &  5.24 &  4.94 & -1.66 &  1.20] \\
		& [5.34 & -0.23 &  3.20 & -0.18 &  3.48 &  1.34]  \\
		\hline
		
		&[5.34 &  6.40 & -1.24 & 5.53 & -2.31 &  2.03] \\
		\makecell{\texttt{ibm\_torino}} & [4.73 & -0.76 & 0.96 & 0.08 & 3.38 & 1.31] \\
		& [4.65 & -0.45 &  1.33 & 0.11 & 3.32 & 1.25] \\ \hline
		
	\end{tabular}
	\label{tab:list_opt_param}
\end{table}

	\end{document}